\documentclass[a4paper,twocolumn]{article}

\usepackage{epsfig}

\begin{document}

\title{Algorithms for Identification and Categorization}
\author{J.M. Cortes$^{1}$\footnote{Present Address: Institute for Adaptive and Neural Computation. School of Informatics. University of Edinburgh. Edinburgh, UK.}, P.L. Garrido$^{1}$, H.J. Kappen$^{2}$, J. Marro$^{1}$, \and C. Morillas$^{3}$, D. Navidad$^{1}$, and J.J. Torres$
^{1}$ \\
$^{1}${\small Institute Carlos I for Theoretical and Computational Physics,
and}\\
{\small Departamento de Electromagnetismo y F\'{\i}sica de la Materia,}\\
{\small University of Granada, 18071--Granada, Spain} \\
$^{2}${\small Department of Biophysics, Radboud University of Nijmegen,}\\
{\small 6525 EZ Nijmegen, The Netherlands } \\
$^{3}${\small Departamento de Arquitectura y Tecnolog\'{\i}a de Computadores,%
}\\
{\small University of Granada, 18071--Granada, Spain.}
}
\maketitle

\vspace{1cm}
\begin{small}{AIP Conference Proceedings 779: 178-184, 2005}\end{small}

\begin{small}{Corresponding author: Joaquin Marro}\end{small}

\begin{small}{mailto:jmarro@ugr.es}\end{small}
\vspace{1cm}

This talk reports on a series of efforts during the last decade aimed at
modeling in a computer the processing of patterns in the brain. In particular,
a main recent interest is to design  fast and reliable algorithms for
\textquotedblleft restoring\textquotedblright\ a pattern, namely,
after degrading highly a pattern, to identify it. The
challenge is to go beyond familiar methods, including Hopfield--like neural
networks \cite{hopfield,amit} which, in spite of a great theoretical
interest, are hampered in practice by the occurrence of mixture states, slow
processing, limited capacity, and inadmissible statistical errors.
Furthermore, we are concerned with an extended definition of
\textquotedblleft pattern\textquotedblright , namely, we plan to deal from a
high--resolution color picture to a complex pattern of behavior or the large
set of qualities which serve to accurately identify a complex company, for
example. This is relevant, in particular, to sociology, the focus in this
meeting. However, the discussion below will avoid any specific application
---except, eventually, for illustrative purposes.

Compared to machines, the human brain performs impressively regarding to 
our recent interest. Unfortunately, the brain strategy in these processes is
not yet well understood. Nevertheless, there are now some indications which
we shall take here as the basic hypothesis to build a model. Firstly, that
processing of patterns is the result of cooperation between many units.
Next, that associative memory in neurobiological systems is the consequence
of, say, an \textit{essentially stochastic} dynamics. That is, taking along
the description to well--defined grounds, dynamics induces stochastic jumps
between \textit{pure} attractors, i.e., fixed points as the ones in the
Hopfield model. Even more, the jumping is often restricted to the elements
in one or a few of the classes in which one may assume the whole set of
attractors to be partitioned. This possibility is interesting because it
implies categorization, namely, classification of the stored patterns
according to some criteria ---e.g., the degree of correlation between the
elements in the class(es). Such behavior has already been reported in
several contexts, including the monkeys cortical activity and the insects
early olfactory processing \cite{monkey}--\cite{olfact}. In fact, there is
some evidence that recognition involves a previous process of class
discrimination (indeed one may recognize a person as familiar and not being
able to recall its identity) \cite{boga}.

The origin for the \textit{essential stochasticity} in actual systems is
likely to be at the synapses, which in practice determine much of the
complex processing of information in the brain \cite{abb}. That is, the
neural activity, which is intrinsically stochastic, further requires the
continuous competition of (fast) \textquotedblleft synaptic
noise\textquotedblright , which in turn is activity dependent, to accomplish
the efficient transmission of information and a variety of computations \cite
{allenPNAS,zadorJN}. Consequently, pursuing recent efforts \cite{bibitchkov}
, we found sensible investigating the effect of stochastic synapses on the
fixed points of the retrieval processes in appropriate attractor neural
networks. With this aim, we believed it convenient to model the synaptic
noise based on the observation that synapses endure short--time \textit{
depression.} That is, periods of elevated presynaptic activity decrease the
neurotransmitter release and, consequently, the postsynaptic response is 
\textit{depressed}, which is likely to positively influence the transmission
of information \cite{abb},\cite{bibitchkov}--\cite{thom}.

We shall briefly illustrate the above ideas in a simple model system which
generalizes several proposals. Consider a set of units (neurons, social
agents, etc.) that change randomly of state with time. Their stochasticity
is regulated by a \textquotedblleft temperature\textquotedblright\
parameter, $T_{0}.$ Each unit is connected to each other with links,
synapses, etc. whose intensities or weights change with time, also at random
but regulated by a different \textquotedblleft temperature\textquotedblright
, $T_{1}.$ This sort of competition was first considered in Refs.\cite{old1}
--\cite{marroB}. To be specific, let $N$ binary (say) neurons with
configurations $\mathbf{S=}\left\{ s_{i}=\pm 1;i=1,\ldots ,N\right\} $
connected by (say) synapses of weight $w_{ij}\in \Re ;$ $i,j=1,\ldots ,N.$
The probability of the state $\mathbf{\Omega }=\left( \mathbf{S},\mathbf{W}
\right) ,$ with $\mathbf{W=}\{w_{ij}\},$ at time $t$ satisfies the familiar
master equation, i.e., $\partial _{t}P\left( \mathbf{\Omega }\right) =\sum_{
\mathbf{\Omega }^{\prime }}\left[ c\left( \mathbf{\Omega }^{\prime
}\rightarrow \mathbf{\Omega }\right) P_{t}\left( \mathbf{\Omega }^{\prime
}\right) \right. -\left. c\left( \mathbf{\Omega }\rightarrow \mathbf{\Omega }
^{\prime }\right) P_{t}\left( \mathbf{\Omega }\right) \right] ,$ with a
transition rate which describes the indicated competition, namely $c\left( \mathbf{\Omega }\rightarrow \mathbf{\Omega }^{\prime }\right) =p\ c^{
\mathbf{W}}\left( \mathbf{S}\rightarrow \mathbf{S}^{\prime }\right) \delta
\left( \mathbf{W}-\mathbf{W}^{\prime }\right) +\left( 1-p\right) \ c^{\mathbf{S}
}\left( \mathbf{W}\rightarrow \mathbf{W}^{\prime }\right) \delta \left( 
\mathbf{S}-\mathbf{S}^{\prime }\right).$ 
The elementary rates $c^{\mathbf{W}}$ and $c^{\mathbf{S}}$ are typically
chosen just by looking for simplicity \cite{marroB}. For instance, one may
take the former as a factorizable function of $2\left( s_{i}/T_{0}\right)
h_{i}^{\mathbf{W}}(\mathbf{S}),$ where $h_{i}^{\mathbf{W}}(\mathbf{S}
)=\sum_{j\neq i}w_{ij}s_{j}$ is the net (pre)synaptic current arriving to
---or local field acting on--- the (postsynaptic) neuron $i.$ Here, $w_{ij}$
is determined in a previous \textit{learning} process, e.g., by the Hebb's
rule \cite{hebb} from $M$ patterns which in this way remain stored in the
system. It is to be remarked that, due to the generality we call for here,
these patterns should be as diverse as to include from a set of orthogonal
or near orthogonal patterns to a set of patterns that are highly correlated
to each other.

\begin{figure}
\centerline{\psfig{file=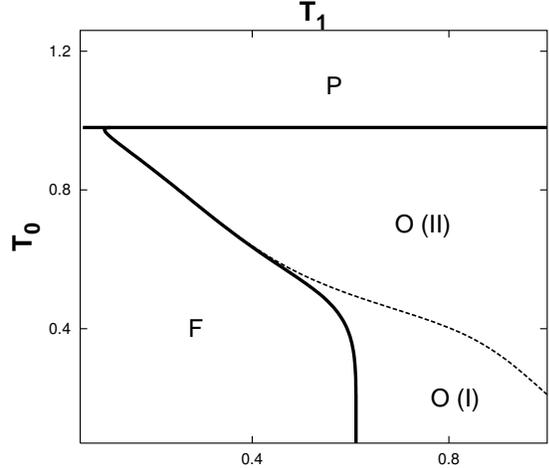,width=7.5cm}}
\caption{ {\small   Parameter
space for $16384$ neurons and $3$ stored, highly--correlated patterns showing $P$
and $F$ phases, as defined in the text, and regions in which the state keeps
jumping between patterns. The jumps are either uncorrelated or
time--correlated for regions O(I) and O(II), respectively. }}
\label{fig1}
\end{figure}

A particularly interesting case occurs for $p\rightarrow 0.$ In fact, this
mimics rapid fluctuations in accordance with the observation of fast noise
mentioned above. Furthermore, the model can then be solved analytically for
some choices of parameters. One is lead in this limit, after rescaling time $
tp\rightarrow t,$ to a situation in which the neurons evolve as in the
presence of a steady distribution for the synaptic fluctuations. The
transition between states is then governed by an effective rate that
involves the noise (effective) distribution. A principal result is that the
ensuing stationary states are out of equilibrium \cite{marroB}. This is
because of the competition between the neural activity and the synaptic
noise, which impedes reaching asymptotically equilibrium. This was first
studied in some detail in Refs.\cite{old2,torresPRA}, and analysis of the
intriguing behavior this may show concerning recognition was first reported
in Refs.\cite{neuronold0,neuronold}.

\begin{figure}
\centerline{\psfig{file=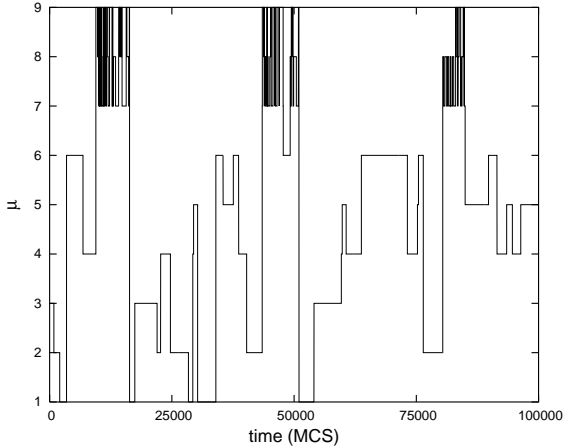,width=7.5cm}}

\caption{ {\small  This shows, as a function of time, the order $ \mu $
 of the pattern which more closely resembles the current state of a
network with parameters in region $O(II)$. This is for $M=9$ correlated patterns
stored in a set of $10^{4}$ fully connected neurons with $T_{0}=0.05$ and $T_{1}=1.25$ times the corresponding critical values. This shows how the neural automata
constantly jumps between the stored patterns. The jumping is apparently
random, but the system detects a family ---consisting of patterns $7$, $8$ and $9$
whose elements are more correlated to each other than with the rest.}}
\label{fig2}
\end{figure}

Also remarkable is that, under appropriate conditions, which include
parallel updating ---as it is usual in cellular automata---, this system
exhibits various retrieval phases. This is illustrated in figure  \ref{fig1}. In addition
to a disordered phase (P) and a phase with associative memory (F), there is
a region in which one observes a continuous stochastic jumping between
attractors (O). Depending on the parameter values, this jumping may be
rather complex \cite{cortes2T}; see  figure \ref{fig2}.  On the other
hand, there is no region in the parameter space with mixture states such as
the ones that hamper recognition in the Hopfield case \cite{amit,neuronold0}
. Also noticeable is that phase transitions may be of first--order, which
implies that the recognition process may occur with negligible error \cite
{neuronold}.

The same model was recently studied to show that the network topology
importantly matters. It is unlikely that natural evolution leads to fully
connected networks, and it was shown \cite{NCtopology} that, in general, the
capacity to store and retrieve is higher for a scale-free topology than for
a comparable highly random-diluted network.

One may explicitly show that some of the jumping phenomena in the model
---e.g., the one shown in figure \ref{fig2} --- is close to the
reported \textquotedblleft familiarity discrimination\textquotedblright\ \cite
{boga} by tuning the model details. For example, one can assume that the synaptic
intensities are now $w_{ij}=\overline{w}_{ij}x_{j},$ where $\overline{w}_{ij}
$ is fixed in a previous \textit{\ learning,} as before, and $x_{j}$ is a
stochastic variable. Let us assume that, in the limit $p\rightarrow 0,$ the
steady noise effective distribution is a product of functions $P(x_{j}|
\mathbf{S})=\zeta \left( \vec{\mathbf{m}}\right) \mathrm{\ }\delta
(x_{j}+\Phi )+\left[ 1-\zeta \left( \vec{\mathbf{m}}\right) \right] \mathrm{
\ }\delta (x_{j}-1).$ Here, $\vec{\mathbf{m}}=\vec{\mathbf{m}}(\mathbf{S})$
is the $M-$dimensional vector of overlaps $m^{\nu }({\mathbf{S}})\equiv 
\frac{1}{N}\sum_{i}s_{i}\xi _{i}^{\nu },$ where $\mathbf{\xi }^{\nu }=\{\xi
_{i}^{\nu }=\pm 1,i=1,\ldots ,N\}$ are the $M$ stored patterns, and $\zeta
\left( \vec{\mathbf{m}}\right) $ stands for a function of $\vec{\mathbf{m}}$
to be determined. This choice aims at modeling the reported short-term
synaptic depression \cite{tsodyksNC,torresNC}. That is, increasing the mean
firing rate results in decreasing the synaptic weight, because the
depression effect here depends on the overlap vector which measures the net
current arriving to postsynaptic neurons.
\begin{figure}
\centerline{
\psfig{file=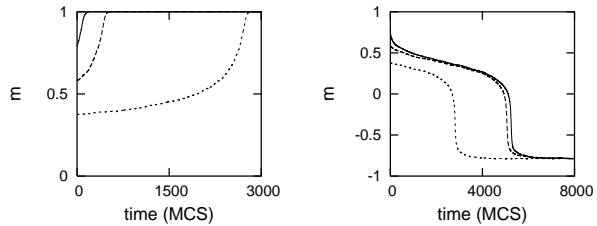,width=8.5cm}
}
\caption{ {\small  Time evolution of the overlap, as
defined in the main text, between the current state and the stored pattern
in Monte Carlo simulations with $3600$ neurons at $T=0.1.$ Each graph
shows different curves corresponding to evolutions starting with different
initial states. The two graphs are for $\delta =0.3$ and 
$\Phi=1$ (right) and $\Phi =-1$ (left), the former corresponding to the Hopfield case lacking the fast noise. This shows the important effect noise has on the network sensitivity to external stimuli. The same behavior occurs as one varies the external
stimulation $\delta$ .}}
\label{fig3}
\end{figure}

The parameter space, i.e., $\left( T,\Phi \right) $ where $\Phi $ is the
parameter that controls now the noise intensity, also depicts the $P$ and $F$
phases, even for the case of sequential updating \cite{new}. The
corresponding transition results of second order only for $\Phi >-4/3,$
which includes the Hopfield classical limit $\Phi =-1$ (lack of depression).
Otherwise, the transition is of first--order ---with excellent recognition
memory due to the sharp behavior just below the transition temperature. More
specifically, there is a tricritical point is at $(T_{c},\Phi _{c})=(1,-4/3).
$ 

\begin{figure}
\centerline{
\psfig{file=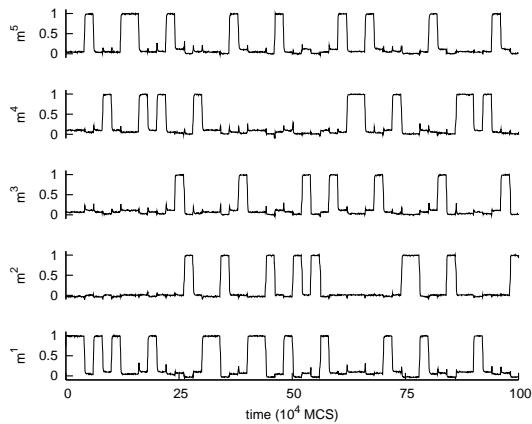,width=7.5cm}
}
\caption{ {\small Time evolution of every overlap during
a Monte Carlo simulation with $400$ neurons, $M=5$ completely uncorrelated patterns, $\Phi=0.05,$ and $T=0.1$. The system in this case was perturbed by the stimulus $-\delta \mathbf{\xi }^{\nu },$ $\delta =0.3,$  with the pattern index $\nu $  changed at random every $20000$ MC steps. The same
experiment with $\Phi =-1;$ i.e., the Hopfield case, produces a constant
signal instead. }}
\label{fig4}
\end{figure}

Also interesting are the consequences of the activity-dependent processes
---which mimic short-term synaptic depression in the model--- on the
retrieval dynamics under external stimulation. That is, one may check the
resulting sensitivity of the network to external inputs. A high degree of
sensibility will facilitate the response to changing stimuli. In fact, this
is an important feature of neurobiological systems which continuously adapt,
and may thus quickly respond to varying stimuli from the environment. A
simple external input may be simulated by adding to each local field a
driving term $-\delta \xi _{i},\ \forall i,$ with $0<\delta \ll 1$ \cite
{bibitchkov}. For a single pattern, $M=1,$ this tries to move the network
activity from the attractor, $\mathbf{\xi ,}$ to the \textquotedblleft
antipattern\textquotedblright , $-\mathbf{\xi }.$ The resulting behavior is
illustrated in figure \ref{fig3} and in the paper \cite{new}.
The graphs in this figure, and similar evidences can be obtained for other parameter values,  clearly demonstrate that presynaptic noise enhances the
network sensitivity to a simple external stimulus. Furthermore, the jumping
phenomena is robust with respect to the type of pattern stored. This is
illustrated by comparison of the situation in figure \ref{fig4} for uncorrelated
patterns, i.e., patterns with mutual overlaps $m^{\nu ,\mu }\equiv
1/N\sum_{i}\xi _{i}^{\nu }\xi _{i}^{\mu }\approx 0,$ and the one in Ref.\cite
{new} for correlated patterns, namely, $m^{\nu ,\mu }=1/3$ for any two of
them. On the contrary, there is no visible structure in the signal response
in the absence of fast noise, e.g., in the Hopfield case as far as $\delta
\ll 1.$ As a matter of fact, the depth of the basins of attraction are large
enough in the Hopfield model, to prevent any jumping phenomena for small $
\delta ,$ except when approaching a critical point ($T_{c}=1$), where
fluctuations diverge. That is, adding fast noise in general seems to
destabilize the fixed point for the interesting case of small $\delta $ far
from criticality.

\begin{figure}
\centerline{
\psfig{file=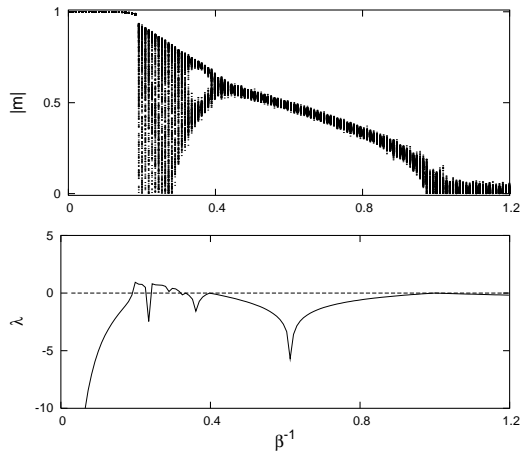,width=7cm}
}
\caption{ {\small The top graph shows, for $M=1$, the absolute value of the overlap as a function of the (neuron)
temperature for  $\Phi =0.05.$ This illustrates, in a MC
simulation with $10^{4}$  neurons, irregular behavior and
a transition at $T_{c}=1$ from the $F$
phase to the $P$ phase as the temperature is increased. The bottom graph shows
the corresponding Lyapunov exponent, as obtained from an accurate
mean--field description. This illustrates eventual eventual chaotic
behavior, i.e., $\lambda >0.$ }}
\label{fig5}
\end{figure}

Summing up, the \textquotedblleft two--temperatures model\textquotedblright\ 
\cite{marroB} may naturally be adapted to incorporate certain
neurobiological observations to obtain an interesting algorithm for memory
recognition, including family discrimination. Using the language of neural
networks, the model has two parameters, say, \textquotedblleft
temperatures\textquotedblright\ which control the stochastic dynamics of
neurons and synaptic weights, respectively. We adapted this to study the
consequences of fast synaptic noise on the attractor of a system with a
finite number of stored patterns. Among the cases already considered, the
most intriguing behavior ensues when the noise depends on the total
presynaptic current arriving to the postsynaptic neuron. This case has been
studied both numerically and analytically by using a mean--field hypothesis.
The numerical work consists of a series of Monte Carlo simulations using
both sequential Glauber, \textit{spin--flip} dynamics and cellular automaton
or Little (parallel updating) dynamics. Most available results concern the
limit $\alpha =M/N\rightarrow 0$ but preliminary Monte Carlo simulations
indicate that our results also hold for a macroscopic number of stored
patterns, $\alpha \neq 0.$

A principal conclusion is that fast presynaptic \textit{noise} may induce a
nonequilibrium condition which results in an important intensification of
the network sensitivity to external stimulation. One explicitly sees that
the noise may turn unstable the \textit{attractor} or fixed point solution
of the retrieval process, and the system then seeks for another attractor.
In particular, one observes jumping from the stored pattern to the
corresponding antipattern for $M=1,$ and jumping between patterns for a
larger number of stored patterns. This behavior is most interesting because
it improves the network ability to detect changing stimuli from the
environment. We observe the jumping to be very sensitive to the forcing
stimulus, but rather independent of the network initial state or the thermal
noise. We also observe that the jumping may become chaotic when the system
is implemented as a cellular automaton with parallel updating. This is
illustrated in figure \ref{fig5}. 

Finally, one may argue that, besides recognition, the processes of class
identification and categorization in nature might follow our model strategy.
That is, different attractors may correspond to different objects, and a
dynamics conveniently perturbed by fast noise may keep visiting the
attractors belonging to a class which is characterized by a certain degree
of correlation between its elements. In fact, as described above, it was
recently reported that a similar mechanism seems at the basis of several
natural phenomena.

We acknowledge financial support from MCyT and FEDER (project No.
BFM2001-2841 and \textit{Ram\'{o}n y Cajal} contract) and from UGR-MADOC agreement.


\begin{thebibliography}{99}
\bibitem{hopfield} J.J. Hopfield, \textit{Proc. Natl. Acad. Sci. USA} 
\textbf{79}, 2554(1982)

\bibitem{amit} D. Amit, H. Gutfreund, and H. Sompolinsky, \textit{Ann. Phys.}
\textbf{173}, 30 (1987)

\bibitem{monkey} M. Abeles, H. Bergman, I. Gat, I. Meilijson, E. Seidelman,
N. Tishby, and E. Vaadia, \textit{Proc. Natl. Acad. Sci.} USA \textbf{92},
8616 (1995)

\bibitem{animals} L.M. Miller and C.E. Schreiner, \textit{J. Neurosci.}
\textbf{20}, 7011 (2000)

\bibitem{scar} S. Scarpetta, L. Zhaopin, and J. Hertz, \textit{Neural Comp.} 
\textbf{14}, 2371 (2002)

\bibitem{oya} T. Oyamada, Y. Kashimori, O. Hoshino, and T. Kambara, \textit{
Biol. Cybern.} \textbf{83}, 21 (2000)

\bibitem{olfact} G. Laurent, M. Stopfer, R. Friedrich, M. Rabinovich, A.
Volkovskii, and H. Abarbanel, \textit{Annu. Rev. Neurosci}. \textbf{24}, 263
(2001)%

\bibitem{boga} R. Bogacz and M.W. Brown, \textit{Hippocampus} \textbf{13},
494 (2003)

\bibitem{abb} L.F. Abbott and W.G. Regehr, \textit{Nature} \textbf{431}, 796
(2004)

\bibitem{allenPNAS} C. Allen and C. Stevens, \textit{Proc. Nat. Acad. Sci. USA} \textbf{91}, 10380 (1994)

\bibitem{zadorJN} A. Zador, \textit{J. Neurophysiol.} \textbf{79}, 1219
(1998)

\bibitem{bibitchkov} D. Bibitchkov, J.M. Herrmann, and T. Geisel, \textit{\
Network: Comput. Neural Syst. } \textbf{13}, 115 (2002)

\bibitem{tsodyksNC} M. Tsodyks, K. Pawelzik, and H. Markram, \textit{Neural
Comp.} \textbf{10}, 821 (1998)

\bibitem{torresNC} L. Pantic, J.J. Torres, and H.J. Kappen, \textit{Neural
Comp.} \textbf{14}, 2903 (2002)

\bibitem{thom} A.M. Thomson, A.P. Bannister, A. Mercer, and O.T. Morris, 
\textit{Philos. Trans. R. Soc. Lond. B Biol. Sci. }\textbf{357}, 1781 (2002)

\bibitem{old1} P.L. Garrido and J. Marro, \textit{J. Phys. A: Math. Gen.}\textbf{25}, 1453 (1992)

\bibitem{old2} P.L. Garrido and J. Marro, \textit{Lecture Notes in Computer
Science} (Springer-Verlag) \textbf{540}, 25 (1992)

\bibitem{torresPRA} J.J. Torres, P.L. Garrido, and J. Marro, \textit{J. Phys. A: Math. Gen.} \textbf{30}, 7801 (1997)

\bibitem{neuronold0} J. Marro, P.L. Garrido, and J.J. Torres, \textit{Phys.
Rev. Lett.}\textbf{\ 81}, 2827 (1998)

\bibitem{neuronold} J. Marro, J.J. Torres, and P.L. Garrido, \textit{J.
Stat. Phys. }\textbf{94}, 837 (1999)

\bibitem{marroB} J. Marro and R. Dickman, \textit{Nonequilibrium Phase
Transitions in Lattice Models}, Cambridge Univ. Press, Cambridge 1999.

\bibitem{hebb} D.O. Hebb, \textit{The Organization of Behavior}, Wiley, New
York 1949.

\bibitem{cortes2T} J.M. Cortes, P.L. Garrido, J. Marro, and J.J. Torres, 
\textit{Neurocomputing} \textbf{58-60}, 67 (2004)

\bibitem{NCtopology} J.J. Torres, M.A. Mu\~{n}oz, J. Marro, and P.L.
Garrido, \textit{Neurocomputing} \textbf{58-60}, 229 (2004)

\bibitem{new} J.M. Cortes, J.J. Torres, J. Marro, P.L. Garrido, and H. J. 
Kappen, \textit{Neural Comp.} \textbf{18}, 614 (2006)

\end{thebibliography}
\end{document}